**Generative deep learning for the inverse design of materials**

Teng Long[1,2,#], Yixuan Zhang[1,#], Hongbin Zhang[1,*]

[1] *Institute of Materials Science, Technische Universität Darmstadt, Darmstadt 64287, Germany*

[2] *School of Materials Science & Engineering, Shandong Universiy, Jinan 250061, China*

*Correspondence to: hongbin.zhang@tu-darmstadt.de*

**In addition to the forward inference of materials properties using machine learning, generative deep learning techniques applied on materials science allow the inverse design of materials, *i.e.*, assessing the composition – processing – (micro-)structure – property relationships in a reversed way. In this review, we focus on the (micro-)structure – property mapping, *i.e.*, crystal structure – intrinsic property and microstructure – extrinsic property, and summarize comprehensively how generative deep learning can be performed. Three key elements, *i.e.*, the construction of latent spaces for both the crystal structures and microstructures, generative learning approaches, and property constraints, are discussed in detail. A perspective is given outlining the challenges of the existing methods in terms of computational resource consumption, data compatibility, and yield of generation.**

## Introduction

Developing and employing advanced structural and functional materials is of significant importance helping us tackle the challenges like energy shortage and information explosion[1]. Conventional materials science research relies mostly on the trial-and-error experiments and individual domain expertise, leading to resource- and time-costly materials development and exploitation. Recently, the data-driven approaches based on machine learning have emerged as the fourth paradigm of materials science, in addition to the approaches based on empirical experimentation, phenomenological theory, numerical simulations using quantum mechanics[2]. For instance, focusing on the core problem of materials science, *i.e.*, to map out the composition – processing – (micro-)structure – property (CPSP) relationships, machine learning can be applied to quantify each link, *e.g.*, identifying synthesis recipe[3], engineering microstructure in additive manufacturing[4], and statistical modelling of various physical properties[5]. However, despite many progresses enabling the acceleration of materials design[2,6–8], the CPSP relationships covering the whole compositional and functional space are still far from quantitatively revealed[9,10], which can be attributed to the high dimensionality of the design space[11]. Therefore, even with the help of machine learning, the current materials design practices are mostly following the many-to-one pattern along the CPSP chain, *i.e.*, predict and synthesize unreported compounds, perform the measurements, and select the one with optimal properties[2,12,13].

To further accelerate materials design, it is essential to access the CPSP chain in a reversed way, *e.g.*, predicting the compounds with target properties together with the processing conditions for the desired microstructure. This leads to the concept of inverse design, which can be carried out based on high-throughput (HTP) combinatorial screening, global optimization, and generative models[14]. Taking the crystal structure – intrinsic property (as given by the crystal structures) as an example, the HTP computational workflow usually consists of three steps: 1) generating new crystal structures by substituting possible atoms of typical prototypes[10]; 2) conducting density functional theory (DFT) computation on the hypothetical structures[15]; 3) screening the compositions with desired properties. In this regard, the HTP method follows again the massive generation and screening scheme, which is still resource-demanding, though it can reduce the cost of materials design by providing guidance to experimental investigations. For solid state materials, HTP screening is performed mostly for the mapping of crystal structure – intrinsic property link based on DFT calculations[16–18], which rely on the implementation of automatized workflows based on platforms such as atomic simulation environment[19], atomate[20], AiiDa[21], etc. Such workflows can nowadays be constructed for many classes of advanced materials such as magnetic materials[22]. It is interesting to get the other simulation tools such as molecular dynamics, phase fields, and finite element (FE) modelling integrated into operative HTP workflows, so that the other links in the CPSP chain can be explored. Furthermore, the global optimization method relies on researchers to extract structural parameters that affect material properties based on existing material structures, and construct reliable surrogate models in order to automatically adjusts these parameters to achieve better properties for the newly discovered materials[23–31]. Such methods include but are not limited to Bayesian optimization and genetic algorithms, which avoid extensive calculation and thus reduce the cost of the design process[32]. An obvious advantage of the global optimization approach formulated based on Bayesian optimization is to implement and execute the so-called adaptive design strategy, which can be directly integrated with experimental investigations and allows an iterative optimization based on active learning[33]. Such a closed-loop adaptive design strategy comprises database curation, surrogate modelling, balanced exploration (searching new regions of the design space) and exploitation (refining known optimal regions), and guided experimental synthesis/characterization, and has been successfully applied to optimize the compositions of NiTi-based shape memory alloys[34], $BaTiO_3$-based piezoelectric ceramics[35], ferroelectric perovskite with higher Curie temperature[36], high entropy alloys with enhanced hardness[37] and as Invar alloys with reduced thermal expansion[38].

In machine learning, the concept of vector space is fundamental and crucial. A vector space is a collection of vectors that can be linearly combined through addition and scalar multiplication. For regression problems in materials science, each vector in the vector space typically represents a sample with multiple features, such as the chemical composition or physical properties of a material. In this space, machine learning algorithms can find a function (i.e., a distribution in terms of statistics) to fit the relationship between sample data and target outcomes[39–41]. However, the actual chemical or physical parameter space is often high-dimensional and complex. Directly learning and optimizing in high-dimensional space can encounter efficiency and accuracy issues. This challenge can be mitigated by introducing the concept of latent space, which is ideally a continuous low-dimensional representation used to describe data in the original high-dimensional parameter space[39,40] and is valuable to better assess the distribution of physical properties. Through an appropriate encoder, each original descriptor vector can be mapped to a unique vector in the latent space; Whereas through a decoder, vectors in the latent space can be decoded back into descriptor vectors with explicit physical meanings. Such a concept of latent space greatly facilitates the applications of Bayesian optimization for materials design, where sampling can be efficiently performed with acquisition functions balancing between exploration and exploitation.

Importantly, with the help of the properly constructed latent spaces, generative deep learning can be valuable to go beyond the known design space to further promote inverse design. On the one hand, there is one unique vector in the latent space corresponding to the original descriptor vector as inputs and hence one specific compound, and all the vectors in the latent space can be decoded into descriptor vectors with explicit physical meaning. On the other hand, such a latent space should be constructed in such a way that the joint distribution $P(x, y)$ of target physical properties $y$ and descriptors $x$ should be faithfully reproduced, which distinguishes it from deterministic models focusing on the conditional distribution $P(y|x)$[11]. Given such a challenge to construct a robust latent space, inverse design based on generative deep learning models has not been extensively done. Furthermore, there are three major machine learning approaches which can be used for generative models, *i.e.*, generative adversarial nets (GAN), variational autoencoder (VAE), and diffusion-based generative model (DGM), including various types such as denoising diffusion probabilistic models (DDPM), denoising score matching (DSM) and discrete denoising diffusion probabilistic model (D3PM) [42–45]. The implementation and application of such models are a bit more technical involved, in comparison to those frequently used forward inference models, such as random forest[46], support vector machine[47], neural networks[48], and transformer[49]. Additionally, another challenging task is how to perform the optimization of physical properties, which is best done in a multi-objective manner and hence entails exploring the usually large chemical/parameter/latent space efficiently to recommend proper candidates fulfilling the requirements[50].

In this review, we focus on the (micro-)structure – property relationships, *i.e.*, crystal structure – intrinsic property and microstructure – extrinsic property (as defined by microstructure), and summarize the current approaches realizing the inverse design of materials based on the generative deep learning. For each section (*i.e.*, crystal structure and microstructure), we elaborate on three critical aspects, namely,

(1) how to achieve efficient communication between machines and humans, *i.e.*, how to generate machine readable descriptors informed with domain expertise,

(2) how to improve the performance of machine learning models,

(3) how to apply constraints so that materials with desired properties will be predicted[11,14,51–56].

Specifically, advances in inverse design are discussed in terms of descriptors or representations for materials, deep learning algorithms, and the integration of property constraints, with related studies summarized in Tables 1 (for crystal structures) and Table 2 (for microstructure), respectively. Finally, the pending challenges will be outlined, together with a bird-view outlook.

## **Section I: Inverse design of crystal structures**

Table 1. List of generative deep learning-based approaches for the inverse design of crystal structures in terms of representation, machine learning model, constraint, reversibility, scope, and efficiency (cf. the main text for details).

| Work | Representations | Models | Constraints | Reversibility | Scope | Efficiency |
|---|---|---|---|---|---|---|
| CrystalGAN[57] | Vector | GAN | None | low | binary and ternary | low |
| IMatGen[58] | Voxels | VAE | Stable or not | high | binary | high |
| CGCNN[59] | Graphs | None | None | None | multicomponent | high |
| iCGCNN[60] | Graphs | None | None | None | multicomponent | high |
| CCDCGAN[61,62] | Voxels | GAN | Formation energy | high | multicomponent | high |
| SmVAE[63] | Graphs | VAE | Four textural properties, three properties related to natural gas separation and three properties | middle | multicomponent | high |
| | | | related to flue gas separation | | | |
| CDVAE[64,65] | Vectors | VAE | Stability | high | multicomponent | high |
| Cond-DFC-VAE[66] | Voxels | VAE | Formation energy, bandgap, bulk/shear modulus, etc. | middle | multicomponent | high |
| FTCP[67,68] | Vectors | VAE | Formation energy, bandgap, Thermoelectric power factor | high | multicomponent | low |
| ZeoGAN[69] | Voxels | GAN | Heat absorption | middle | ternary | high |
| Composition Conditioned Crystal GAN[70] | Vectors | GAN | Pourbaix stability and the band gaps | high | ternary | low |
| PGCGM[71,72] | Vectors | GAN | Formation energy | high | multicomponent | high |
| DP-CDVAE[73] | Vectors | DDPM & VAE | None | high | multicomponent | high |
| Scalable Diffusion using UniMat[74] | Vectors | DDPM | Composition | high | multicomponent | high |
| MatterGen[75] | Vectors | DDPM & DSM & D3PM | Composition, symmetry, magnetic, electronic, mechanical properties and supply-chain-risk | high | multicomponent | high |

The generative deep learning-based inverse design method actively optimizes the non-linear relationship between material structures and properties without external intervention, thus attracting intensive attention [51,76,77]. In general, such a method extracts knowledge from the existing structure-property datasets and applies the learned knowledge in designing new materials. One of the main challenges of applying the inverse design method for crystal structures is the descriptors of crystalline materials. In Table 1, we summarize the state-of-the-art crystal structure generative models based on the representations of crystal structures, the model architectures, whether there are constraints, the reversibility of the used representations, the application scope of the models, and their generation efficiency.

Representations, also known as descriptors or features, are a set of parameters that represent the structural features of materials. In this section, they refer to the parameters corresponding to the crystal structures that are used as input for the machine learning models. It is noted that crystalline parameters as specified in standard .cif files are often not sufficient for machine learning because the three-dimensional (3D) periodicity is not automatically encoded, and also that the crystal structures in real space are bidirectionally mapped to the descriptors in the latent space.

Models are deep learning generative models used to predict new crystal structures. These models are a type of machine learning model that learns to generate new data similar to the data used for training. From the mathematical perspective, these models attempt to mimic the distribution of the training data through various approaches and metrics.

The term *constraint* comes from nonlinear programming problems in operations research, referring to additional conditions that must be satisfied while achieving the main objective. In inverse design, the main objective is the generation of new crystal structures, whereas desired intrinsic properties can be considered as constraints when generating crystal structures. A constraint that can be integrated into the objective function or as a tolerance for screening after the generation process.

Reversibility refers to the ability to achieve bidirectional mapping between crystal structures in real space and descriptors in latent space. Specifically, it includes two aspects (1) whether the model can correctly recognize the known crystal structures, *i.e.*, uniquely encoding and decoding the existing crystal structures, and (2) whether the newly designed crystal structure can be correctly recognized, *i.e.*, sampling the latent space for reasonable descriptors which can be decoded to crystal structures in the real space. Therefore, this indicator is generally characterized by the reproducibility rate of the lattice parameters and atomic positions in the crystal structures during the transformation between real and latent spaces.

The scope indicates the compositional space covered by the inverse design model, e.g., binary/ternary means that the model can only be used in a specific binary/ternary system (sometimes transferrable for other binary/ternary compositions), whereas a multicomponent model can generate new crystal structures with more than three elements.

Efficiency refers to the ability to generate new materials effectively, *i.e.*, the ratio of the number of crystal structures fulfilling the constraints with respect to the total number of generated structures via decoding.

### Crystal structure descriptors

In order to enable the inverse design of crystalline materials, it is necessary to create descriptors mapping the crystal structures for the deep learning models[78,79]. Descriptors of crystalline materials can be divided into three categories: 1) elemental descriptors that describe the chemical information of materials constituents[80–82], 2) structural descriptors that describe the geometric information of crystal structures[83,84], and 3) combined descriptors describing both the chemical and structural information[59,85]. Either the elemental descriptors or the structural descriptors highlight only partial information of crystalline materials, which are not sufficient to specify the crystalline materials. Even though it is possible to combine them, the heterogeneity and poor connectivity make the description of crystalline materials inefficient. In comparison, the combined descriptors integrate the essential information of crystalline material as a whole and therefore have the potential to reconstruct the crystalline materials. However, not all combined descriptors can be used in the deep learning methods, as inverse design methods also require that identifiable crystalline materials should be extracted via decoding from the corresponding descriptor[86]. Since all crystalline materials are a periodic arrangement of atoms in the 3D space, the unit cells are good starting points for generating descriptors[87]. In general, a unit cell is composed of lattice parameters, atomic coordinates, and atomic species. The lattice parameters describe the size and shape of the unit cell, the atomic coordinates indicate the atomic position in the unit cell, and the atomic species mark the elements at the corresponding positions. Thus, the descriptors of the crystalline material used in generative models should allow the extraction of all such information to reconstruct the unit cell. There are deep learning models focusing on the chemical information[88–90] and its extracting[91,92], the structural information is indispensable. In this regard, generative deep learning to predict crystal structures remains a significant challenge[61].

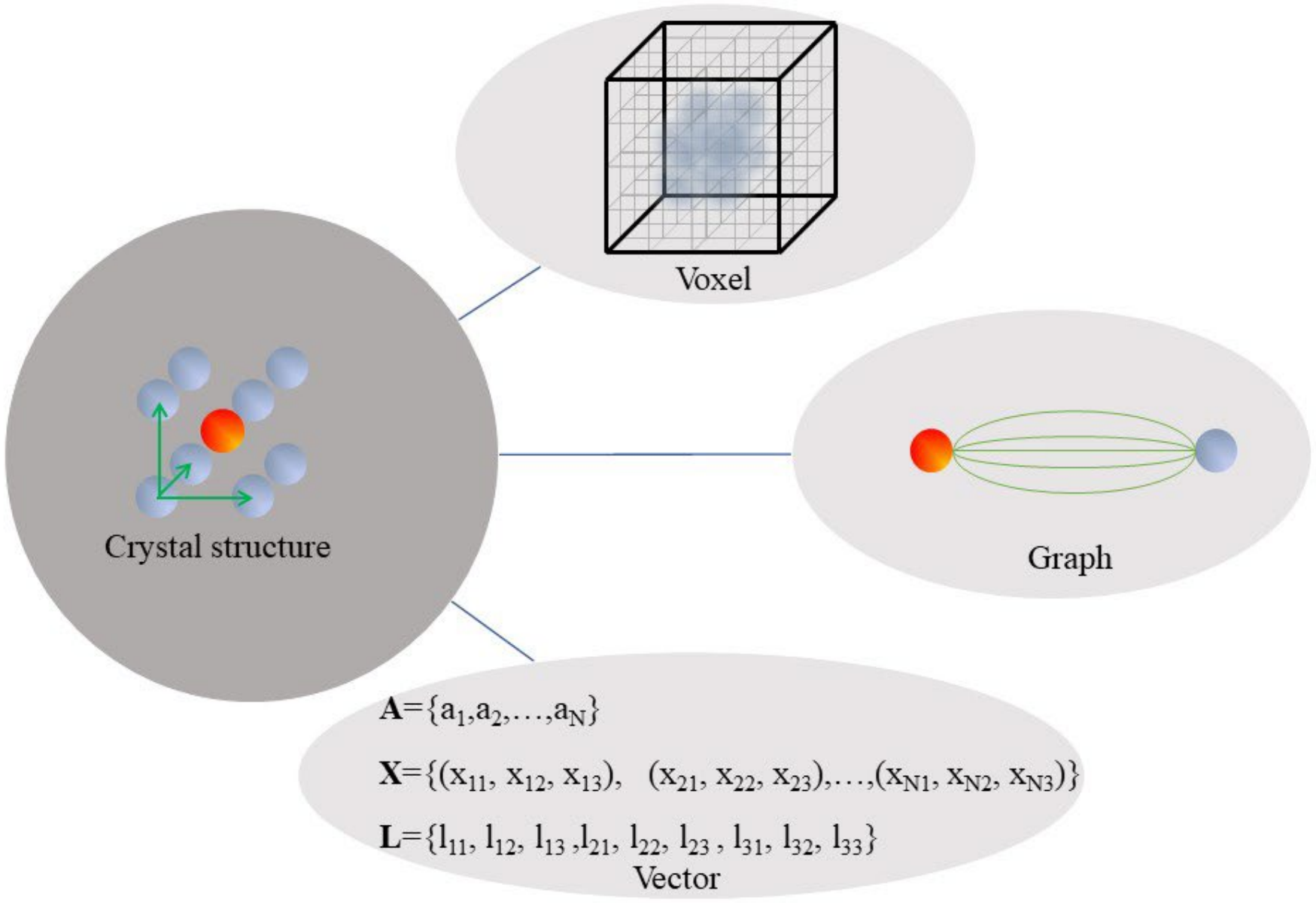

**Fig. 1.** Common descriptors of crystal structures, with the crystal structure on the left and the voxel descriptors, graph descriptors and vector descriptors from top to bottom on the right.

Nevertheless, there are a few existing solutions which have been applied successfully for generative deep learning, including the voxel method[58], the vector method[64,65,73–75], and the graph theory method[63], as demonstrated in Fig. 1. The voxel method uses 3D grids to voxelize the crystal structures, i.e., creating 3D voxels to record the atomic species at the corresponding atomic positions[86], and either the voxels themselves or the vectors obtained by encoding the voxels are used to get the latent space[93–95]. Due to usually required dense grids for voxelization and comparatively limited number of atoms in unit cells, the voxels are significantly sparse (if only the atomic positions are labeled). Thus, severe information loss will occur during the reconstruction process, i.e., the positions of the atoms cannot be reproduced accurately. In order to avoid sparsity, atomic positions can be transformed into atomic densities by introducing a Gaussian smearing with element types being scale parameters. Such a representation scheme can achieve a reconstruction ratio of about 70%, making it an acceptable solution for generative deep learning of crystal structures.

Subsequently, Noh et al. proposed an improvement to this method[58], with three essential modifications: 1) the lattice parameters are also treated using voxels, i.e., one extra voxelization explicitly created to store the shape and size of the unit cell, dubbed as the cell voxel hereafter, 2) The scale parameters for Gaussian smearning are considered as hyper-parameters, instead of using the element types, and 3) individual voxels are created for different elements (called the basis voxels) and stored in different channels. The improved voxel method was successfully applied for the binary V-O system, with a reconstruction ratio of more than 90%. The voxel approach was extended to multicomponent systems by Court et al. [66] by abandoning using multiple channels to store elemental information and resolving back to describing different elemental species by scale parameters, while retaining the lattice parameter voxelization. Such a scheme enables the design of multicomponent crystal structures[66] with the cost of a reduced reconstruction ratio. Long et al. further improved the voxel approach by decomposing different elements into different channels, fixing the scale parameters, and introducing empty channels to maintain the consistency of the voxel data, resulting in an overall reconstruction ratio above 85%[62]. It is noted that Kim et al. successfully used the voxel method to describe porous materials, demonstrating the potential of the voxel approach for microstructure representation[69].

The voxel approach has become one of the most successful solutions in crystal structure representation, with a high reconstruction ratio and the potential to extend to microstructure. However, it exhibits a few shortcomings: 1) it does not reflect the symmetry information, which is vital in crystallography, 2) it cannot identify the similarity of the structures after rotation, *i.e.*, it cannot ensure the rotational symmetry of the descriptors, and 3) it requires a large number of computational resources for data processing.

The graph theory approach uses vertices and edges to describe atoms and interatomic connectivity, as well as channels to reflect the elemental information with weighted based on the interconnections between the individual atoms[60,96,97]. In this regard, it accounts for the translational and rotational symmetry properly. For example, the properties of the corresponding atoms and bonds are added by the channels in crystal graphs, giving rise to the necessary description of the crystal structures. The graph-based descriptors have become one of the most attractive methods[59]. Xie et al. used the one-hot features to store information for both vertices and edges and found that such features suffered from poor reversibility to reconstruct the crystal structures. This dramatically limits the application of the graph theory approach in generative deep learning models. Even though the method has been further developed by Park et al., there are still no reports about reconstructing crystal structures from graphs[60,96,97]. Yao et al. have solved the problem from a different perspective, i.e., using vertices to represent a group of atoms in the graph and edges to represent the interactions between groups[63]. This approach enables the description of massive systems with the cost of relying on other computational methods to achieve reconstruction. The method has shown good performance in the design of materials such as MOF. Therefore, we believe the graph theory approach can accurately capture the symmetry and periodicity of crystalline materials, but its poor reversibility limits its applications in generative models.

The vector method uses vectors to directly describe the atomic species and positions and the lattice parameters. Kim et al. proposed a method to incorporate symmetry and robustness through data augmentation, such as adding perturbations and rotations to the original data during the training process, and successfully applied it for the Mg-Mn-O system[70]. However, this did not solve the problem of the low reconstruction ratio of the vector method. Given that the atomic position information is complicated to reconstruct accurately, Ren et al. enhanced the description of atomic positions by describing the structures of crystalline materials using both Cartesian and Fourier transformed coordinates[67]. After encoding and decoding, such two types of information can be cross-validated, thus significantly improving the reconstruction ratio of the crystal structures. However, the generation process has extremely low efficiency due to the redundant representation. In contrast, Xie et al. proposed a solution by using elemental species, atomic coordinates, and lattice constants as descriptors. Assuming that the lattice parameters can be accurately reconstructed, the scale of the noise of elemental species and atomic coordinates can be measured by scoring neural networks with data argumentation[64]. As a result, a more accurate reconstruction of elemental species and atomic coordinates can be achieved. At the same time, they proposed a pre-optimization scheme to determine the corresponding atomic coordinates and elemental species during the generation process, which guaranteed the stability of the generated structures. Furthermore, Zhao et al. improved the validity of the generation by introducing the element properties, atomic pairwise distance constraints and structural symmetry into the generation model, resulting in a higher success rate and more symmetrized structures in generation[71,72]. Recently, Yang et al. have defined a four-dimensional scalable vectorized material representation UniMat based on the periodic table of elements utilizing the a priori knowledge of the periodic table[74], with also good performance for both reconstruction and generation. In short, the vector method turns out to be a good solution for describing crystalline materials and becomes extremely promising after the solution proposed by Xie et al. In particular, the rise of diffusion generative models (DGMs) based on Markov diffusion processes can improve the reconstruction rate significantly, which guarantee robust forward and backward mappings and hence reliable reconstruction and generation of crystal structures.

**Crystal structure generative models**

Generative models are the key to inverse design, in order to learn existing crystal structures and design new ones. Three typical generative deep learning models, namely, variational autoencoder (VAE), generative adversarial network (GAN), and diffusion model, are illustrated in Fig. 2.

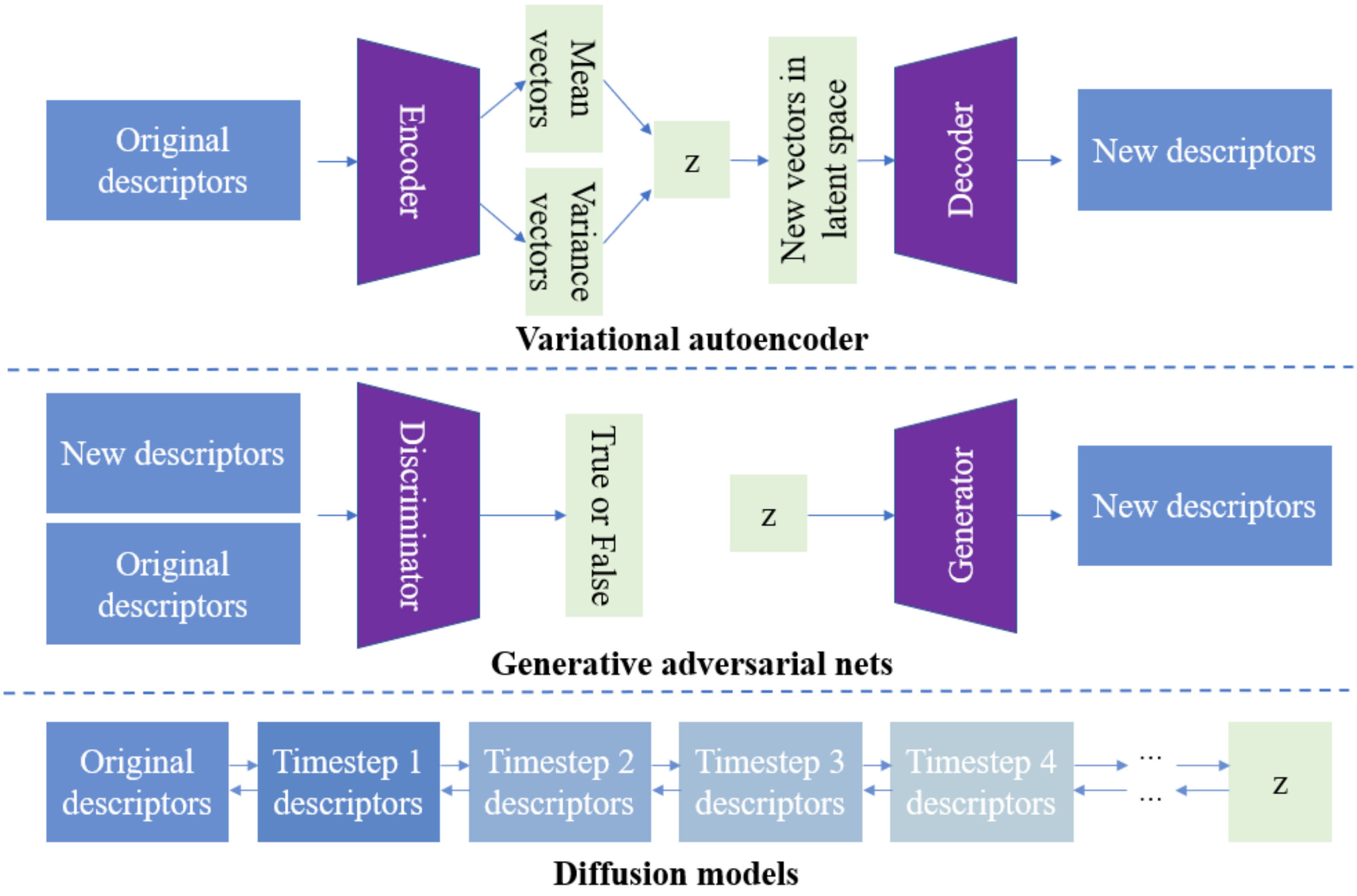

**Fig. 2.** The architectures of commonly used generative deep learning models, *e.g*., VAE, GAN, and diffusion model, where z denotes the latent space.

The VAE model consists of two parts, i.e., the encoder and the decoder. The former takes descriptors as discussed above as inputs and maps them into vectors in the latent space. The vectors in the latent space should satisfy a specific distribution, and they contain all necessary information to reconstruct the inputs. The decoder uses the latent space vectors as input to perform the reconstruction. After training the VAE model, new vectors are randomly sampled in the latent space according to the learned distribution, hence generating new crystal structures[42]. Decoders and encoders are generally neural network models, specifically common types are convolutional neural networks (CNNs)[98] and graphical neural networks (GNNs)[99].

Gomez-Bombarelli et al. applied the VAE model to design molecules[100]. The first successful application of the VAE model to crystal structure design was conducted by Noh et al.[58], where a 2D CNN-based VAE model was used to design new V-O materials with high efficiency. It was demonstrated that the VAE model is more efficient in generating new V-O materials than the global optimization model. The VAE model was subsequently used by Court et al. to generate Heusler alloys, chalcogenides, and binary alloys with various compositions[66]. They adopted the idea of using voxel descriptors, but they constructed a 3D CNN model directly on the voxels without using crystal images[77]. However, this study also pointed out the shortcomings of the VAE model, i.e., the generated structures were mostly not stable and needed to be re-optimized by further DFT calculations.

Since descriptors based on the vector method can be applied to better specify the crystal structures, Ren et al. applied the VAE model with vector descriptors using CNN constructs[68]. Due to a large amount of noise in the vector descriptors generated with the VAE method, only a few portion of the generated descriptors can be transformed into reasonable crystal structures[101]. The efficiency of the VAE model in generating new materials is therefore low, with a success rate of less than 1%[102]. Xie et al. proposed the crystal diffusion VAE (CDVAE) model to address this problem[64]. They assumed that there was a deviation of the structures generated by the VAE model from their stable state, so the feasibility of finding the target materials could be increased by analyzing the noise distribution[103–105]. Such analysis relies on the noise conditional score network (NCSN) model. Besides, CDVAE also takes the stability of the materials into account by exhausting the possibilities of the generated materials, so only optimal structures can be selected[103]. The CDVAE model also demonstrates the possibility of using GNN constructions in the VAE model.

In general, the VAE model can generate complex images and control the hidden space distribution. It is believed that the choice of distribution functions integrates the domain knowledge on the crystalline materials, which is the key to accelerate model training[52]. However, the choice of the distribution functions in VAE significantly affects the model performance, which may lead to a significant drop in the performance of such models.

Conceptually, the GAN model consists of a generator and a discriminator where the generator aims to generate new descriptors from a set of random arrays, while the discriminator will determine whether the distribution of the generated descriptors is statistically consistent with the descriptors of the known crystal material[43,106]. Thus, the generator tries to ensure that the generated descriptors cannot be identified by the discriminator as generated descriptors by improving the similarity to the existing structures, while the discriminator will be trained to find the difference between the generated and the actual descriptors. The generator and the discriminator are trained to compete against each other simultaneously, and eventually, the generator will be capable of generating descriptors of crystalline materials that are sufficiently good.

The GAN model was first applied to generate crystal structures in the CrystalGAN model by Nouira et al.[57]. They used vectors to describe the structures of binary materials, i.e., A-H and B-H, and used the GAN model to generate ternary structures, A-B-H. Although the generation ratio demonstrated in this work was not high, CrystalGAN successfully generated Ni-Pd-H crystal structures using Ni-H and Pd-H crystal structures as inputs, enabling the generation of heterogeneous crystal structures. Subsequently, Long et al.

used 2-dimensional crystal images based on voxelization as inputs and outputs of a GAN model[61]. This method was successful when acting on a specific binary system, Bi-Se, and was able to find new Bi-Se structures that did not exist in the training set, with a decent generation ratio. Kim et al. successfully generated new structures of Mg-Mn-O using the GAN method with higher generation rates compared to the VAE models. This demonstrates the advantages of the GAN method in exploration[70]. Long et al. then embarked on an attempt to use the GAN model for multicomponent systems, successfully generating crystalline materials with various compositions[62]. Nevertheless, the generation ratio dropped in its application of multicomponent systems, mainly due to the limited number of crystal structures in the training dataset for a vast space of compositions.

The GAN model has demonstrated advantages in generating new structures using image-based descriptors, and also has the potential to be combined with vector-based descriptors (no research has been reported). It is worth noting that GAN models learn the distribution of the latent space by themselves implicitly and therefore are not immune to mode collapse. This leads to a decrease in the structural generation ratio of such models.

Diffusion models, which are inspired by non-equilibrium thermodynamics, use denoising to generate data. The process of learning by denoising involves two Markov Chain processes: the forward process and the reverse process. During the forward process, random noise is gradually added to the data at a series of time steps from $t_1$ to $t_n$, with samples at the current time step drawn from a Gaussian distribution conditioned on the samples at the previous time step, and the variance of the distribution following a predetermined schedule. After long enough forward time steps, the samples become standard Gaussian distributions. In the reverse process, starting from a standard Gaussian distribution, the noise is reduced at each time step in the backward direction from $t_n$ to $t_1$[44,107]. After training, we can use the DGM to generate data by simply passing randomly sampled noise through the learned denoising process. In fact, the Markov chain process in the algorithm shifts the goal from matching distributions in the data space and the potential space to finding a strategy that describes the direction in which the noisy samples at the current time step are most likely to change towards a steady state within every time step, *i.e.*, what the DGM actually does can be seen as finding a hypothetical potential energy surface which, by inputting a noisy sample and the corresponding timestep, will output a more stabilized state at the next timestep.

Xie et al. employed the score-based network NCSN as a decoder to denoise the perturbed structures to the stable crystal structures[64,65]. Pakornchote et al. added an extra DDPM model on the top of CDVAE, where CDVAE is used to predict the lattice parameters and the number of atoms in the unit cell, and the DDPM model helps to denoise the fractional coordinates and predict the atomic coordinates[73]. Yang et al. directly applied the DDPM on their UniMat representation for both unconditional and conditional crystal structure generation[74]. Based on the CDVAE descriptors, Xie et al. applied the D3PM, DSM and DDPM with limit distribution for atom type denoising, fractional coordinate denoising and lattice distribution denoising, respectively, where the matching of these three denoising models is by score evaluation using an SE(3)-equivariant GNN named GemNet-dT[75].

The current trends show that crystal structure generation is a complex problem and requires multiple models acting together. In addition, the consideration of whether a crystalline material can be synthesized is an interesting direction, as it has been mostly discussed for molecules[108,109]. Currently, crystal structure generation models are mostly CNN-/GNN-based VAE and GAN models, but diffusion models are also starting to enter the mainstream. Nevertheless, if transformers can be applied to construct generative deep learning models, their capability will possibly be further enhanced[110,111].

**Crystal structure constraints**

While current generative deep learning models have been mostly applied on generating new crystal structures, integrating the physical property prediction in such models has started to attract more and more attention in recent years. In this regard, the property constraints can be incorporated either as a screening approach or as an optimizing approach, as shown in Fig. 3. The former involves two steps: 1) generation of the new structures and 2) screening structures with desired properties. Noh et al. applied this approach by firstly generating many $V_xO_y$ structures using the VAE model and then selecting those with good thermodynamic stability with formation energies below -0.5 eV/atom[58]. Such an approach allows for exploring a larger chemical space, but it cannot be applied to optimize the target properties during the design process, which dramatically slows down the design of functional crystalline materials. This problem was ameliorated in the work by Court et al.[66], where they generated a large number of crystal structures using the VAE model and used a forward prediction model, i.e., the CGCNN model developed by Xie et al. to predict the properties (including bandgap, mechanical properties, and formation energy) of the generated structures. Such a scheme improved the efficiency of screening functional materials by selecting only crystal structures exhibiting target properties in the generation process[59], without considering the material properties during the training process though.

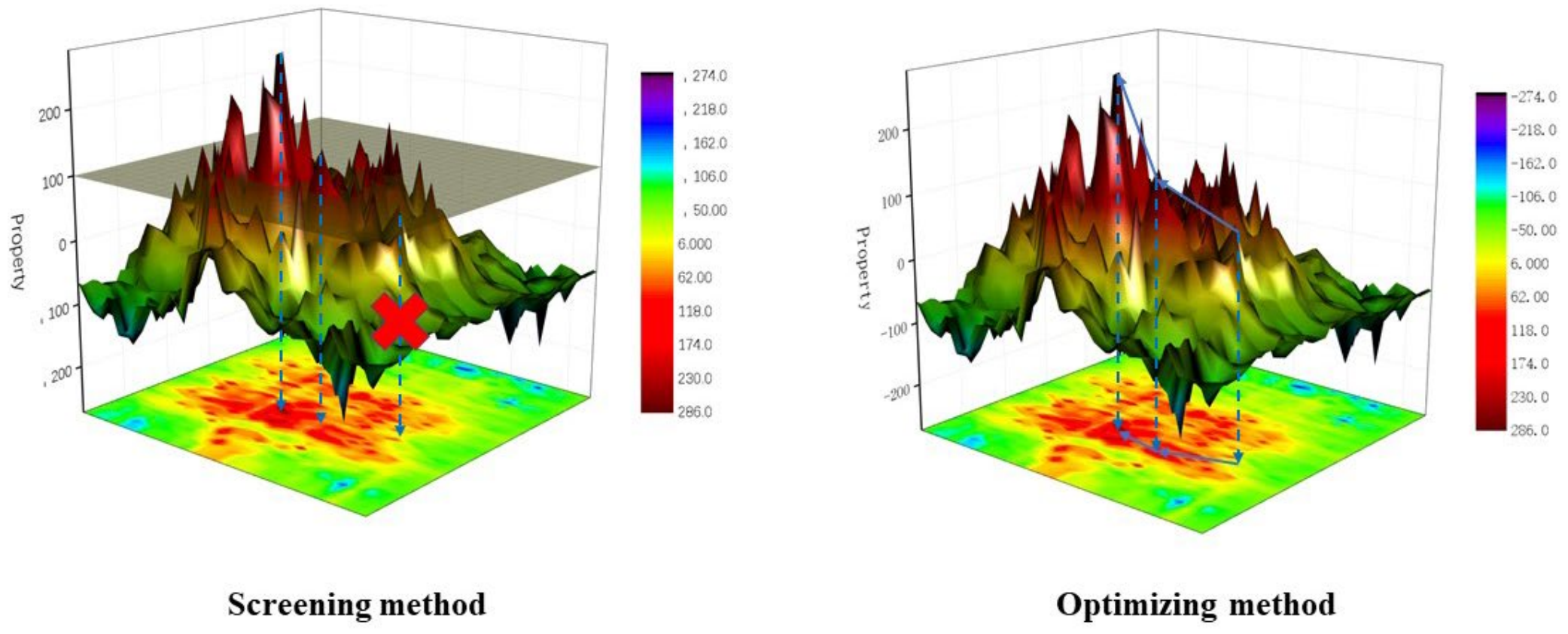

**Fig. 3.** Two methods of adding constraints to the generative models: the screening method (left panel) and the optimizing method (right panel).

To address such an issue, Long et al. added material properties in the loss function of the GAN model as a back propagator to achieve joint optimization of material properties and crystal structures[61]. It is noted that modifying the loss function will affect the discriminator's evaluation of the generated structures. For instance, if the formation energies of the generated structures are too high, the value of the loss function will increase and therefore prevents the generation of such structures. Detailed analysis reveals that this approach improves the efficiency of generating the structures with target properties by 100 times, which can significantly improve the generation ratio of functional crystalline materials[62]. Kim et al. added heat absorption as a constraint in their GAN model to design porous materials, enabling the prediction of porous crystalline materials with heat absorption in a specific range[69]. Yao et al. also added other functional properties to the loss function in the VAE model, leading to the direct design of functional MOF materials[63].

Taking this a step further, Xie et al. introduced the process of structure optimization directly into the generative model. By generating lattice parameters, atomic positions, and element species individually and exhausting the possibilities based on the above constraints, they could predict materials with desired properties[64]. Unlike the cases of GAN and VAE, in diffusion models, the cross-attention layers enable the incorporation of constraints into the joint distribution, thus controlling the direction of structure generation towards distributions with target properties. The corresponding control nets can be added during the initial model training process as done in UniMat[74], or later to fine-tune the pre-trained model as implemented in Mattergen[75].

In short, the essence of property constraints is to balance exploration and exploitation. Unconstrained models can be considered as explorative models, focusing on the design of new crystal structures in the latent space that are distinct from the existing ones[112]. In contrast, constrained models explore the latent space to identify crystal structures with desired properties, emphasizing the design of novel structures based on the available information. According to our research on adding constraints, the balance of exploration and exploitation will significantly affect the performance of generative models, which could be a future research direction for generative deep learning models.

**Section II: Inverse design of microstructure**

Table 2. List of deep learning-based microstructure generation methods, their features, advantages and disadvantages.

| Work | Models | Constraints(properties) | Database | Dimension | Comments |
|---|---|---|---|---|---|
| Yang, et al.[113] | GP-Hedge + GAN | Style loss<br>Optical property | Synthetic microstructure images by GRF (5000). | 2D | GP-Hedge as constrain for effective property searching |
| Iyer et al. [114] | ACWGAN-GP | different cooling methods | UHCSDB (172). | 2D | Investigated the processing-structure linkage |
| Lambard et al.[115] | StyleGAN2 with ADA | None | Ferrite-martensite SEM (3000). | 2D | Low data requirement |
| Chun et al.[116] | GAN | None | HMX SEM image (1). | 2D | Seamlessly generation across dimensions |
| Ma et al.[117] | pg-GAN<br>Pix2Pix GAN | Different processing methods | U-10Mo SEM-BSE (272). | 2D | Spatial exploration by GAN needs careful evaluation |
| Fokina et al.[118] | StyleGAN | None | SEM of Alporas aluminum foam (1);<br>Micro-CT of Berea sandstone (1); | 2D | multiple image resolutions |
| | | | Micro-CT of Ketton limestone (1). | | |
| Squires et al.[119] | DCGAN | None | FIB-SEM of a three-phase solid oxide fuel cell anode (1);<br>hypoeutectoid steel image from DoITPoMS (1). | 2D | Using microstructural inpainting to recover the defects and unwanted artefacts |
| Lee et al.[120] | DCGAN<br>Cycle GAN<br>Pix2Pix | None | OM (9216) and SEM (10000) images of steel surface;<br>OM (3045) and SEM (3186) images with x2000 magnification of steel surface;<br>SEM of Li-battery cathode/anode with spherical (3045) and wiry (3186) morphologies;<br>Hand-drawn sketches (1200) and martensitic (440) micrographs. | 2D | Using large amount of dataset<br>Image style changing |
| Kench et al.[121] | SliceGAN | None | Micrographs meet special criteria from DoITPoMS (87) | 2D→3D | Construct 3D microstructure from 2D images. |
| Hsu et al.[122] | Wasserstein-GAN | None | SOFC anode microstructure PFIB-SEM Synthetic 3D (1). | 3D | Generated structures outperform Dream.3D results in electrochemical simulations. |
| Henkes et al.[123] | Wasserstein-GAN | None | Synthetic spherical inclusions microstructures;<br>Micro-CT scanned wood-plastic composite microstructure (1). | 3D | Investigate the influence of network topology, filter number, and geometrical and physical inductive biases. |
| Gayon-Lombardo et al.[124] | DC-GAN | None | Li-ion battery cathode XCT (1); SOFC anode XCT (1). | 3D | Generate arbitrarily large synthetic microstructural volumes and the periodic boundaries. |
| Cang et al.[125] | Constrained-VAE | Style loss | Sandstone microstructures (200). | 2D | Incorporate a style loss as constrain into the model training. |
| Kim et al.[126] | GPR-VAE | Ultimate tensile strength; uniform elongation toughness | Synthetic dual-phase steels (4000). | 2D | Investigate the structure-property linkages in continues space. |
| Düreth et al.[127] | Diffusion model | Microstructure class | NFFA-Europe, UHCSDB (13000); Fiber Composite (36). | 2D | High model stability |
| Azqadan et al.[128] | Diffusion model | Processing parameters | AZ80 magnesium alloy components with different casting-forgings (27) | 2D | Investigate the process-structure linkage |
| Lyu et al.[129] | Diffusion model | Permeability | A 2D database comprising various types of structural features (8000) and 3D porous material database (3000) | 2D & 3D | Achieve 3D microstructure generation with property control |
| Lee et al.[130] | Diffusion model | Volume fraction Effective elastic modulus Light intensity change ratio | Micro-CT images of SAOED composites (300) and stress fields calculated by FEM (1920) | 2D | Employ multiple target material properties into conditioned DGM |

Microstructure is a general term that describes the structural features and topological arrangements at a particular length-scale. e.g., nanoscale or macroscale. That is, it refers to the arrangement of a material's constituent components, including atoms, molecules, grains, and phases, at a microscopic level, which can be formulated as the size, shape, orientation, and distribution of microstructural units including grains, phase boundaries, defects, inclusions, and precipitates. Microstructure plays a crucial role in determining the extrinsic properties of materials, such as their strength, ductility, toughness, hardness, conductivity, magnetization, and corrosion resistance. For example, a material's strength depends on the size and distribution of the grains, while its thermal conductivity depends on the shape and orientation of the grains., Therefore, understanding the morphology of microstructure is critical for designing new materials with desired properties.

Microstructure morphology can be analyzed using both experimental and computational methods. These methods help to identify key features of the microstructure, which can be used to investigate material properties and design new materials with desirable properties. Experimental techniques used for microstructure characterization include optical microscopy, scanning electron microscopy (SEM), transmission electron microscopy (TEM), X-ray diffraction (XRD), electron backscatter diffraction (EBSD), atomic force microscopy (AFM), X-ray computer tomography (XCT), micro computed tomography (micro-CT) scans, and Four-dimensional scanning transmission electron microscopy (4D-STEM), etc. These techniques produce 2D or 3D microstructure images with varying levels of resolution and depth, as summarized in Table 3 (see [131,132] for more details). Computational methods can be used to further analyze microstructure morphology, including quantification and reconstruction of statistically equivalent microstructures. The features considered in such theoretical methods can be classified into several categories, including statistical functions (e.g., two-point correlation function, linear-path function, and two-point cluster correlation function), deterministic physical descriptors (e.g., volume fraction, total surface area, and number of clusters), statistical physical descriptors (e.g., cluster's nearest center distance and orientation angle of a cluster's principle axis), spectral density function, texture synthesis and multiple-point statistics[133].

Once the explicit or implicit features of microstructure have been characterized, it is possible to reconstruct the microstructure using different methods. For example, if statistical descriptors are used for the characterization, reconstruction can be achieved through optimization using methods such as simulated annealing (SA), genetic algorithms, or other gradient-based algorithms. Alternatively, reconstruction of texture synthesis or multiple-point statistics can be done through Gaussian pyramids[133]. There are also established software tools, such as Dream3D[134] and OptiMic[135], which incorporate these statistical and computational algorithms for simulating microstructure. These tools are designed to facilitate the reconstruction process and provide a platform for analyzing and designing materials with specific microstructures and properties.

While statistical and physical features can be useful for characterizing microstructure morphology, they are limited to specific microstructure systems or morphologies. Additionally, they may overlook important features or relationships within the microstructure, such as small-scale variations or complex interdependencies. Machine learning generative models offer a more comprehensive and flexible approach to microstructure representation and design. These models can learn from large datasets of microstructures and generate new ones that conform to specific constraints or objectives. Moreover, they can uncover complex relationships between the microstructure features and material properties, enabling the design of microstructures with optimized and novel properties. In recent years, generative deep learning models have shown a great promise in advancing microstructural materials science, and are becoming increasingly important for designing new microstructures with tailored properties. Therefore, the development and application of generative deep learning models for microstructure design are a critical area of research that will have significant impact on a wide range of technological fields. In Table 2, we summarize the state-of-the-art microstructure generative models in terms of the representations of microstructure (mainly the data types and feature scales used by the model), the model architectures, the constraints applied during the generation, and the accuracy and interpretability of the models.

Table 3. The experimental measuring approaches of microstructures.

| Experimental Approach | Resolution (typical) | Data Type | Merits | Drawbacks | Comments |
|---|---|---|---|---|---|
| Optical microscopy | Micrometers | 2D | Simple, widely used | Low resolution | Useful for observing overall morphology and microstructure |
| Scanning Electron Microscopy (SEM) | Nanometers | 2D | High resolution, surface imaging, compositional analysis | Requires vacuum environment, sample preparation | Widely used in materials science |
| Transmission Electron Microscopy (TEM) | Nanometers | 2D | High resolution, internal structure and composition, crystal defects | Requires thin samples, complex sample preparation | Useful for studying microstructure and composition at the nanoscale |
| X-ray diffraction (XRD) | Angstroms | 2D | Determines crystal structure and composition, non-destructive | limited to crystalline materials | Useful for identifying and characterizing different materials and their properties |
| Atomic Force Microscopy (AFM) | Nanometers | 2D/3D | High resolution, surface topography and roughness, mechanical properties, non-destructive, operates in air and liquids | Slow scanning | Useful for studying surface properties and topography at the nanoscale |
| Electron Backscatter Diffraction (EBSD) | Nanometers | 2D | Determines crystallographic orientation, non-destructive | Surface analysis only | Useful for studying crystallographic orientation and deformation history |
| X-ray computer tomography (XCT) | Micrometers | 3D | High resolution 3D imaging, internal structure, non-destructive | Requires high dose of X-ray radiation | Useful for imaging biological and porous materials |
| Micro Computed Tomography (micro-CT) scan | Micrometers | 3D | High resolution 3D imaging, internal structure | Requires X-ray radiation, limited to relatively small samples | Useful for imaging biological and porous materials |
| Four-Dimensional Scanning Transmission Electron Microscopy (4D-STEM) | Nanometers | 2D | High resolution imaging, electron diffraction | Requires complex data analysis, limited temporal resolution | Useful for studying dynamics at the nanoscale |

**Microstructure representations**

In contrast to the representation of crystal structures, which only requires a quantitative description of atomic positions, atomic species, and crystal lattices, the construction of microstructure descriptors is impeded by the numerous metrics that entail intricate intercorrelations with the desired properties and the vast visual phase space of microstructures. As microstructural morphology becomes more complex, traditional microstructure representations such as volume fractions, Minkowski tensors and spatial correlations [133,136] become less effective. Contemporary microstructure generation tends to train a machine learning model directly using real or synthetic microstructure images. However, the preparation of microstructure samples is complex and expensive, the existing experimental images are insufficient to comprehensively cover the vast visual phase space. Consequently, a complete construction of microstructure features as microstructure descriptors is currently not feasible[137]. At present, the assessment of microstructural features is largely based on human intuition and varies depending on the specific system being analyzed and the method of measurement employed. However, with the advent of computer vision (CV) technology, it has become possible to quantitatively detect and analyze abstract information within images[138,139], CV algorithms can be employed to digitize experimental microstructure images and then processed to achieve various research goals. Consequently, utilizing experimental or synthetic digitized microstructure images directly to describe the microstructures is a more feasible and dependable approach compared to constructing statistical microstructure descriptors. This approach addresses the first issue mentioned earlier in this paragraph. In order to gain a better understanding of the visual phase space of microstructures, CV algorithms can be combined with unsupervised generative machine learning models, as discussed above for crystal structure generative deep learning models. The resulting algorithms can extract abstract features from digital crystal images, comprehending the distribution of microstructures, and such generative deep learning models with the learned distributions can be applied to generate new microstructure images in the reversed way with desired physical properties optimized.

To build a high-quality uniform generative model from scratch, a large and diverse dataset with over 50,000 samples is essential. Although established microstructure image databases, such as the NFFA-EUROPE SEM Dataset (~21169), ASM Micrograph Database (~4100), UHCSDB[123] (~961) and DoITPoMS(~818) provide tens of thousands of high-quality microstructure images and are wildly used for benchmarking new analysis techniques or training machine learning models, the current scale of existing micrograph datasets falls short of the threshold required for a uniform generative model. Therefore, proper pre-processing and data enrichment are vital as they help machine learning models capture and understand the underlying features in the images. Due to the limitations of experimental approaches, there is a deficiency of intermediate data that covers the vast visual phase space. As a result, the current generative model tends to focus on local features by segmenting a large image into small ones or on a small region of the visual phase space with a specialized dataset. To choose between a local generative model trained on specific experimental micrographs or a uniform generative model trained on simulated synthetic augmented micrographs, one must weigh the benefits and trade-offs. The localized generative model provides a precise description of the local data distribution, leading to an accurate understanding of the microstructure-property relationships in the specific system. This approach can help to uncover system-specific mechanisms and theories. In contrast, the uniform generative model provides more opportunities to explore unexplored regions in the visual phase space, increasing the probability of finding potential new materials with desired properties. The following paragraphs will describe specific examples or applications of each approach, combining the summarized information in Table 2 to illustrate their benefits and trade-offs further.

Firstly, for the generative models that focus on local features, the key characteristic is that only a few microstructures (less than 10) exist in the database. The main idea behind local feature generation models is to use convolutional layers to capture the underlying statistical features of the images and then reconstruct or generate the microstructures for specific use. It is worth mentioning that the reconstructed microstructures may differ from the original structures, but the morphological statistical distribution remains the same. Therefore, these models are mainly used for constructing larger microstructures for simulation purposes or recovering defects or blurred regions caused by experiments.

Ref. [116,118,119] constructed generation models by randomly sampling regions in a single image, focusing solely on the local features of a specific material compound. These models are primarily aimed at microstructure modeling for simulation calculations or image inpainting to recover blurred regions. Particularly, Chun et al. used a size 3000×3000 pixels SEM image of a class V cyclotetramethylene-tetranitramine pressed energetic material, and generated a training dataset containing 12500 images with a size of 161×161 by cropping the original image in random position. Fokina et al. tested two systems during their investigation, the SEM image of Alporas aluminum foam and the digital rock by micro-CT. For Alporas aluminum foam, 16000 size 128×128 images were random cut from the original 751×751 SEM images; Whereas for the digital rock, 10000 400×400 2D slices of Berea sandstone and 10240 256×256 2D slices of Ketton limestone were randomly cut from the original micro-CT images separately. Squires et al. used FIB-SEM of a three-phase solid oxide fuel cell anode and a hypoeutectoid steel image from DoITPoMS as their benchmark tests for image inpainting, an occluded region was first set and in each training iteration a batch of training images were sampled from the unoccluded region.

Ref. [121–124] also focused on capturing local features but in 3D reconstruction. The same strategy of conducting random segmentations on microstructure was employed to build the training database. The segmented 2D or 3D samples were used to train the generative models, which were then utilized for 3D structure reconstruction for simulation purposes. Specifically, Kench et al.[121] selected 87 out of 818 2D micrographs in the DoITPoMS dataset based on a set of exclusion criteria, and the inpainting technic was applied to remove the scale bars in the images, to maintain the data features to the great extent under the extremely data-scarce situation. Hsu et al.[122] trained their 3D Wasserstein-GAN (WGAN) model using the 3D data of the solid oxide fuel cell (SOFC) anode containing yttria-stabilized zirconia, nickel, and pore phases measured by Xe plasma focused ion beam combined with SEM (Xe PHIB-SEM), with the volume of 110×124×8 $\mu m^3$, during the training a subvolume of 65×65×65 $nm^3$were randomly sampled. Henkes et al. [123] used two datasets to train and test their model performance, one was the dataset of synthetic spherical inclusions microstructure with $32^3$ voxels, another was a micro-CT scanned wood-plastic composite microstructure of 950×240×850 voxels, and then 1000 samples of 64×64×64 voxels were picked out via Latin-Hypercube sampling for model training. Gayon-Lombardo et al.[124] also used two datasets for model benchmarking, the first was a Li-ion battery cathode measured by XCT with a size of 100.7×100.3×100.3 $\mu m^3$, the second one is the same SOFC anode dataset as in Ref.[122]. During the training of each dataset, more than 10000 subvolumes were extracted using an 8-voxel-stride sampling function. It is noteworthy that all current generative models for 3D microstructures focus on local features. This could be attributed to the challenges posed by the scarcity of 3D data and the significant information demand required for the reconstruction process during the model's training.

Secondly, for generative models that focus on small regions of visual phase space, the dataset size lies between hundreds to thousands across several morphologies of a specific compound system. These models can well describe the local morphological distribution of the system and thus often have good interpretive ability in describing the process-structure-property linkage.

Among the papers listed in Table 2, Ref. [115,117,120] investigated the performance of generative models in describing the local morphological distribution of microstructures. Lambard et al.[115] used a datasets of 3000 SEM images of 30 ferrite-martensite dual-phase steels with different martensite fractions to train a StyleGAN2 model. Ma et al. [117] used 272 SEM-BSE images of depleted U-10Mo alloy from 10 classes, each class denoted a different processing history. The original images were cropped into 10080 512×512 pixels images with the augmentation including horizontal shift, rotations and horizontal/vertical flipping. The classes information was used for training a random forest (RF) classification model. Lee et al.[120] collected four distinct datasets to evaluate the performance of different GAN models in microstructure generation, including one large dataset containing 19216 256×256 pixels steel micrographs with different magnifications (9216 from OM and 10000 from SEM), one dataset containing 6231 256×256 pixels steel micrographs with ×2000 magnifications (3186 from OM and 3045 from SEM), one dataset of 971 spherical and 1130 wiry SEM micrographs of size 128×128 pixels, and one dataset containing 1200 hand-drawn sketches of steel SEM images and another 440 martensite steel SEM images.

Ref. [113,114,125,126] took a step further, they used constrained model to investigate these local regions where more promising properties are expected. Yang et al.[113] used a 5000 synthetic micrograph database of size 128×128 pixels generated by GRF model, where the parameters in GRF model were carefully controlled to guarantee the dispersity of the data. Iyer et al. [114] cropped the original 172 steel SEM images in UHCSDB into 7000 size 128×128 pixels micrographs. Cang et al.[125] used a small dataset of 200 sandstone micrographs and considered the properties including Young's modulus, diffusivity and fluid permeability to test the performance of their morphology constrained VAE model. Kim et al. [122] used 4000 statistically synthetic dual-phase steel micrographs for their GPR-VAE model performance benchmark. Azqadan et al. [128] utilized 27 AZ80 magnesium alloy components with different casting-forgings to train a DDPM conditioned on different processing parameters. Lee et al. [130] used 300 SAOED composites Micro-CT images to train a unconditioned DGM and generated 5760 random samples. The target properties of 1920 out of 5760 samples were calculated using FE modeling and were used to train a U-Net surrogate model. The target properties of the rest samples were predicted by this U-net model, where the conditioned DGM model was trained based on such a dataset with 5760 samples.

Finally, for uniform generative models, besides the requirement of large datasets, it is also essential to have a stable model to ensure the reliable understanding of global data distribution. In Düreth et al.'s [127] work, they trained their state-of-the-art DDPM on a large dataset consisting of more than 13000 raw data in 13 different classes, sourced from NFFA-Europe and UHCSDB. Lyu et al. [129] used a 2D database comprising various types of structural features consists of 8000 micrographs and 3D porous material database consisting 3000 microstructures to train their class-conditioned DDPM. These large and diverse datasets on the one hand ensured good coverage of the visual phase space and on the other hand met the large data requirements due to the huge scale and complex diffusion process of the DDPM model.

**Microstructure generative models**

According to the analysis above about using VAE and GAN models for the inverse design of crystalline materials, it is clearly that researchers have roughly the same preference for VAE or GAN models. However, the situation changes dramatically for microstructure generative deep learning models. As shown in Table 2, there are only 2 out of 15 investigated papers using VAE models. This is partially because the generated images from VAE are always blurred than those obtained using GAN. Such an issue can be straightforwardly solved in crystalline material inverse design via an extra procedure, for example, DFT or molecular dynamic structural relaxations which can be done in a HTP way. Similar extra treatments are challenging for microstructure, because there are no reliable physical simulations available which can reproduce experimentally available microstructures. Nevertheless, to understand the reason for the blurry generations of VAE, we need to start with the essential difference between the VAE and GAN models in terms of how they describe their data density function.

VAE is designed to learn an explicit density function in the latent space, which serves as an effective feature representation to approximate the true distribution in addition to enabling the generation of new data. That is, VAE also expects to model the data distribution explicitly and obtains the well-defined latent representation which can be used for inference. This is achieved by optimizing the loss function of VAE, which involves maximizing the evidence lower bound of likelihood. The optimization process ensures that the encoded posterior distribution of latent vectors, conditioned on the current samples, is as close as possible to the normal distribution with the help of reparameterization (calculated by the Kullback–Leibler (KL) divergence), and that the decoded results from these vectors with added noises are as close as possible to the original data. However, VAE assumes that all the data obeys a multivariate Gaussian distribution with independent components, which may not hold for real arbitrarily complex distributions. Consequently, this assumption can lead to suboptimal sample quality in the latent space and correspondingly blurred reconstructed results in the physical space.

In Table 2, investigations that utilized VAE were primarily interested in its capacity to construct effective latent representations. Cang et al.[125] introduced a morphology constraint during the training of the VAE model by implementing an additional morphology style loss penalty generated by a pre-trained VGG net. Despite the VGG model being trained on image training sets rather than VAE latent representations, the structural-morphology distribution learned by VGG prevented the mismatching of neighboring latent samples to significantly deviated images. Besides, to prevent the model from generating a cluster distribution, a model collapse loss was also incorporated. Furthermore, the added losses resulted in a reduction in the weight of the KL divergence loss in the total loss, indicating a decrease in the uncertainty considered during training. These two factors combined to guarantee a higher efficiency of training and higher clarity of the micrograph generation of this constrained VAE than normal VAE. Kim et al.[122] employed Gaussian process regression (GPR) to depict the associations between the latent space vectors and mechanical properties. Additionally, they utilized the trained GPR to identify the highest uncertainty points, which served two purposes: firstly to enhance the diversity of the dataset, and secondly to swiftly identify the microstructures with desired target properties.

In contrast to VAE, GANs do not operate using an explicit distribution function. Instead, they learned to generate samples from the training distribution through a zero-sum game between the generator and discriminator. This gives rise to two fundamental differences between GAN and VAE. Firstly, there is no uncertainty in the mapping from the latent space to the samples, and secondly, the inference using latent space representation is problematic. Furthermore, optimizing the zero-sum game using Jensen-Shannon (JS) divergence can lead to instability during the optimization process and often ends up in saddle points. Various attempts have been proposed to address this instability by modifying the model structure, latent space and loss function. This trend can also be seen implicitly in the development of microstructural GAN models.

In the case of 2D microstructural GAN models, Ref. [113,114,116,119] modified the loss function and latent space of the GAN model to tackle their specific problems. Yang et al.[113] incorporated the same style loss and model collapse loss used in Cang et al.[125] into the GAN model, and they also introduced a GP-Hedge Bayesian optimization (BO) to control the sampling in latent space in order to generate microstructures with more promising optical properties. Iyer et al. [114] proposed an auxiliary classifier Wasserstein GAN with gradient penalty (ACWGAN-GP) model for synthesizing steel microstructures under specific processing conditions. To achieve this, they trained an additional classifier alongside the GAN model to ensure the model could differentiate between different processing conditions. As a result, the classifier loss and a gradient penalty loss (to prevent model from collapse during training) were incorporated into the loss function, and an extra processing condition vector was introduced into the latent space. Chun et al.[116] employed the modified latent space and loss function to a simple GAN model, they achieved the morphology generating control of HMX microstructures by extending the loss of GAN into patch-based loss, and applying a combined latent vector (the global morphology parameter sampled from an uniform distribution and the local stochasticity parameter sampled from an uniform random distribution) of each grid. These modifications resulted in a scalable mapping of each grid to some specific region within the model perception in the micrograph, consequently enabled tractable control of the microstructure morphology generation. Squires et al.[119] proposed a deep convolutional GAN (DCGAN) model for the inpainting tasks of recovering defects and unwanted artifacts in micrographs via generator optimization approach or seed optimization approach. To address the requirement of matching boundaries during inpainting, for the generator optimization approach, a content loss function for annulus region was included in the loss function; And for the seed optimization approach, an extra seed optimization was carried out separately when generating new micrographs.

The remaining 2D microstructural generation models primarily addressed their target problems through modifications to the model structure. For example, Fokina et al.[118] utilized a StyleGAN architecture in combination with image quilting technique for their microstructure reconstruction, which resulted in the generation of high-resolution and high-quality micrographs. In the StyleGAN model, a multi-layer perceptron (MLP) was used to learn an affine transform to project the latent vectors into an intermediate latent space that is disentangled from the data distribution. This transform was then applied to each convolutional block through a normalization algorithm called adaptive instance normalization to achieve better control over the generation process. Similarly, Lambard et al.[115] employed an updated StyleGAN2 architecture on a small dataset consists of 3000 dual-phase steel micrographs. An adaptive discriminator augmentation (ADA) was also implemented to stabilize the training of GAN model within limited data regime. This allowed for the generation of microstructures with good quality and good interpolations between microstructures. Ma et al.[117] tested the performance and interpretability of two different GAN models: the progressively growing GAN (pg-GAN) and the pix2pix GAN. The pg-GAN model trains progressively from low to high resolution data layer by layer, using reliable weights obtained from previous layer as a weighted residual during the training of the next layer. The pix2pix GAN uses a labeled image as a constraint to a U-net type generator to achieve point-to-point mapping. Finally, Lee et al.[120] employed DCGAN for generating new virtual micrographs, resulting in highly realistic and visually appealing microstructures. They also utilized Cycle GAN and Pix2Pix GAN to conduct style transform between OM and SEM or sketches and SEM images.

In 3D microstructure generation, due to the further reduction of data amount compared to that of 2D microstructure, the aims were mainly focused on reconstructing larger 3D microstructure from the data of a single microstructure, where only the local features were required. Therefore, the original GAN architecture was sufficient for handling such tasks. For example, Gayon-Lombardo et al.[124] applied a DCGAN in their study, and Hsu et al.[122] and Henkes et al.[123] used the WGAN with the Wasserstein distance as the loss function to guarantee the smooth gradient during the model training. They also explored the performance of an S4 equivariant network that considers rotational equivariance. While the S4 CNN showed better quality in generating microstructure, it required a significant amount of computational time and memory.

As we can see from previous part, generative models have advanced considerably in the last few years, with efforts focused either on improving the performance of VAE within reasonable computational limits via better variational posteriors, or on improving the stability of GAN through better loss functions and discriminators. However, a fundamental question arises: Can we reap the benefits of both VAE and GAN, rather than being forced to choose between Scylla and Charybdis? Specifically, is it possible to create a generative deep learning model that trains a simple objective function and is compatible with highly expressive neural networks? The DDPM model proposed by Ref. [44,107] modified the goal of generator from 'mapping standard Gaussian distributions to data distributions' to 'fitting the inverse process of a defined Gaussian Markov Chain which maps standard Gaussian distributions to data distributions'. In this way, the generator only needs to match each small step of the inverse process corresponding to the forward Markov Chain, rather than optimizing the generator and the variational posterior/discriminator simultaneously. The work in Ref. [127–130] has demonstrated the great potential of DDPM on capturing complex microstructural morphologies and controllable new micrograph generations.

**Microstructure constraints**

Microstructure plays a critical role in determining the mechanical, thermal, electrical, and magnetic properties of materials. The morphology properties of microstructures, such as grain size, shape, orientation, distribution, and phase composition, as well as porosity, connectivity, and tortuosity of the pore network, which have significant influence on material properties such as strength, ductility, electrical and magnetic behavior, can be characterized using a variety of experimental techniques. For instance, the mechanical properties of microstructures, such as strength, ductility, toughness, and fatigue resistance, can be characterized using techniques like nanoindentation, tensile testing, and fatigue testing. The thermal properties of microstructures, such as thermal conductivity, specific heat capacity, and thermal expansion coefficient, can be characterized using techniques such as laser flash analysis, differential scanning calorimetry (DSC), and thermomechanical analysis (TMA). The electrical properties of microstructures, such as electrical conductivity, resistivity, and dielectric constant, can be characterized using techniques such as impedance spectroscopy and dielectric spectroscopy. The magnetic properties of microstructures, such as magnetization, coercivity, and remanence, can be characterized using magnetic measurements techniques, such as vibrating sample magnetometry (VSM) and magnetic force microscopy (MFM). Understanding the relationship between microstructure and material properties is crucial for optimizing the performance of materials in various applications, including energy storage, catalysis, and biomaterials.

In addition to experimental methods, simulation methods can also be used to evaluate extrinsic properties driven by microstructure, usually multiscale simulations in order to capture the underlying mechanisms and to bridge to device performance. For instance, FE analysis is a widely used simulation method for evaluating mechanical properties of microstructures, such as stress distribution and deformation behavior. Molecular dynamics (MD) simulations can be used to study thermal and mechanical properties of microstructures at the atomistic level, providing insights into properties such as thermal conductivity and strength. Phase-field simulations can be applied to model the evolution of microstructure during various processing routes, such as solidification or annealing, and predict the resulting microstructure properties. Monte Carlo simulations can be used to simulate the behavior of a system with randomly distributed variables, such as the distribution of pores in a microstructure, and predict the resulting properties. These simulation methods are increasingly being used in conjunction with experimental techniques to gain a more comprehensive understanding of extrinsic properties and their relationship to microstructure.

Despite the plethora of experimental and simulation methods available for evaluating such microstructure-derived extrinsic properties, there is still a lack of a comprehensive process-microstructure-property database comparable to those available for crystalline materials, such as the Materials Project or the Inorganic Crystal Structure Database (ICSD). The diversity and complexity of microstructures, as well as the vast number of morphologies that can influence their properties, make it challenging to develop a complete and accurate database. As a consequence, apart from the screening method or optimizing method like crystal constrains, microstructure generative models also developed active learning type constrains on the top of these two methods to generate new microstructures with desired properties. The development of such microstructure constraint generative models represents a promising avenue for advancing the field of microstructure design and understanding the relationships between microstructure and extrinsic properties.

For microstructure-property constraints, Cang et al.[125] added a style penalty evaluated by a pre-trained VGG net and Gram matrices to the VAE. It is showed that adding a meaningful additional loss can greatly improve the performance of the VAE, providing theoretical support for the application of physical constraints in generative models. Yang et al.[113] transplanted the same type loss function to a GAN model and added GP-Hedge evaluation to actively generate new structures with desired optical properties. Kim et al.[126] used the same GPR active learning strategy on a VAE model to establish the relationship between microstructure and mechanical properties. Lyu et al.[129] used a class-conditioned DDPM to generate porous 3D microstructures with different permeabilities. Lee et al.[130] embedded volume fraction, effective elastic modulus and light intensity change ratio as three controlling properties into conditional DGM for the inverse design of mechanoluminescence particle composites.

In addition to the microstructure-property linkage, the processing-microstructure mapping can also be used as a constraint to the microstructure generative model. This relationship describes how the microstructure is influenced by the processing history, such as the thermal treatment conditions. By incorporating such processing information into generative models, a more accurate prediction of the resulting microstructure and its properties can be achieved. This can be done by treating the processing conditions as different classes to the microstructures and introducing a classifier into the generative model. The relationship between the processing conditions and resulting microstructures can be learned from a large dataset collected through either experimental or multi-physics simulation approaches, such as FE analysis or MD. The resulting generative model can then be used to predict new microstructures under different processing conditions and to optimize processing conditions for optimal microstructures.

There have been various attempts to incorporating processing-structure constraints into generative models. For screening-type constraints, Ma et al.[117] used a random forest classifier to predict processing conditions and evaluate microstructure representations. In contrast, Iyer et al. [114] incorporated processing constraints in an optimizing way, predicting different cooling methods using an additional classifier built into the discriminator and introducing corresponding scoring loss and latent vectors in the GAN architecture. The work of Düreth et al. [114] found that DDPM could generate different classes of microstructures by comparing descriptors such as spatial three-point correlations or Gram matrices, without the need for an explicit classifier in the DDPM model. Azqadan et al. [128] applied the same class-conditioned DDPM structure to AZ80 magnesium alloy to achieve the processing parameter control without any explicit classifier. This is because the MC process used in DDPM naturally ensures a healthy and traceable distribution mapping, which provides good classification ability. These methods enable microstructure generation under different processing conditions and complete major parts in the closed-loop of CPSP inverse design, providing a more comprehensive understanding of the mechanisms behind microstructure formation and its impact on material properties.

**Section III: Challenges and Outlook**

Given the strong interest of the materials science and relevant communities and significant progresses on data-driven acceleration of materials design, the inverse design strategy offers an opportunity to tackle the CPSP relationships in an innovative way, boosted by the application of generative deep learning and Bayesian optimization. However, there are still many challenges to be properly handled.

For the crystal structure inverse design, the low success rate in designing crystalline materials can be attributed to several factors. Firstly, the exploration space of materials that humans can investigate is limited. Traditional experimental methods for discovering new materials are time-consuming and resource-costly, thereby restricting the number of materials that can be explored. Additionally, the complex nature of crystal structures makes it challenging to synthesize new materials even with the help of automated inverse design methods. Secondly, an essential challenge in crystal structure inverse design is the development of effective descriptors. The existing descriptors may not capture all the critical information needed, *e.g.*, how to properly describe and impose the crystallographic symmetries including the point group and the translational symmetry (i.e., 3D periodicity) during the crystal structure generation. Thirdly, the efficiency of generative models is a critical factor in crystal structure inverse design. Generative models must be able to sample from a vast space of possible crystal structures, identify promising candidates with desired properties, and fine tune themselves with the examined new structures efficiently. Therefore, the updating of generative models that can better describe the latent-real space distribution is of vital importance.

Similar challenges hold for the microstructure inverse design such as limited exploration space and high demand for efficient models. Although the use of synthetic micrographs can help to expand the database and overcome the shortage of data, the subtle differences between the real and synthetic microstructures and the complex parameter settings during the machine learning simulations can lead to biased distributions. From the physical point of view, the formation and evolution of various microstructures is driven by the thermodynamic and diffusive processes depending on the initial compositions and heat treatments. That is, it is critical to evaluate the composition – processing – (micro-)structure links in the CPSP chain by either systematic experimental investigations or quantitative simulations of such processes, *e.g.*, via phase field modelling[140]. This entails the establishment of an inverse design paradigm for the composition – processing – (micro-)structure links. Such a paradigm is valuable so that extra physical constraints can be applied in the generative models, giving rise to better efficiency when dealing with deficient datasets. On the other hand, it makes a lot of sense to curate digitized microstructure database following the FAIR principles[141], so that generative models can be developed to minimize the differences between the real and synthetic microstructures.

Thus, to address the challenges in the development of generative deep learning models for both crystal structures and microstructures, it is crucial to improve the quality and diversity of the available data. While experimental data collection remains an essential approach, the high resource- and time-cost involved in syntheses and measurements make it impractical to rely solely on the experiments to increase the amount of data. As an alternative, one possible solution is to establish a synthetic-real data linkage using style-transfer or multi-fidelity ML models. This approach allows for the modelling of more real structures using synthetic ones, or the simulation of more properties using real structures. Additionally, data mining of previous results using generative pre-trained models, such as GPT4 [142], has proven to be a useful tool in collecting and analyzing data and images from papers and technical reports, leading to the development of a homogeneous and extensive database. Ultimately, improving the quality and diversity of data will enhance the performance and accuracy of generative models, allowing for more efficient and effective applications in crystal and microstructure inverse design. But in contrast to fields like images or natural language, where vast amounts of data are generated daily on the internet, the natural sciences often struggle with limited data availability, even with the aid of the previously mentioned data augmentation methods. Therefore, in the short term, relying solely on enhancing the dataset to improve the expressive power of structural generation models is not a practical solution. Instead, we may need to explore better self-supervised learning methods to overcome this challenge.

In terms of descriptors and representations, crystal structures present challenges due to their large number of atoms and lattice sizes, which exceed the limits for voxel-based descriptors. Including more materials without reducing reconstruction accuracy can lead to memory issues. Vector-based descriptors also face challenges due to their inherent heterogeneity, requiring significant zero padding. Additionally, the meaningful atomic positions due to the periodicity in crystal structures make vector-based descriptors difficult to use. The graph theory methods have not fully addressed the challenge of reconstructing crystal structures, but optimization during reconstruction may be a possible solution. For both types of generative models, physics-informed descriptors that map between the structures and properties are crucial for success. These descriptors should be explainable and understandable, providing insights into the underlying physical mechanisms governing processing-structure-property relationships.

From the modelling perspective, the learning speed of generative models, being VAE or GAN, is not yet fast enough. In order to improve the efficiency of generative model structures, it is necessary to incorporate more physics. Quantum machine learning shows promising acceleration effects for unsupervised learning model training, which may help to solve the problem of slow training [143]. The diffusion models simplify the training objective function, make the generative models much more stable and compatible with highly expressive neural networks. In addition to the advancements in model structures, current transformer architectures have demonstrated promising results in terms of enhancing data comprehension. For instance, the Graphormer [144] applies the transformer architecture to the GNN to tackle over-smoothing problems, leading to a significant improvement in the GNN's expressive ability. Therefore, there is still a considerable scope for further improvements through the combination and reconfiguration of model architectures. Furthermore, multiscale models that can connect crystal structures and microstructures can provide a more comprehensive understanding of the processing-structure-property relationships. One possible approach to achieving this is to use the weight from the crystal generative model as additional information in the low-depth layer of the microstructure generative model.

All of the aforementioned approaches provide an opportunity to develop a large, general model similar to GPT and generalize all tasks within this model with the assistance of domain knowledge models. Furthermore, with a quantitative mapping of the CPSP relationships, generative deep learning models can make efficient and reliable predictions, resulting in an effective surrogate model which can be combined with the experimental investigations to further improve the closed-loop adaptive design approach. In particular, the resulting inverse design strategy is promising for the realization of future autonomous experimentation for inorganic solid state materials[145], which will significantly accelerate the materials discoveries and utilizations. Such a promising paradigm involves leveraging a generic generative model to predict possible new materials with desired properties and the processing conditions to synthesize and optimize the microstructure, as well as using robotic systems to conduct high-throughput experimentation, enabling the rapid exploration of a vast search space of synthesis conditions and material compositions. By analyzing the experimental data, the generative model can be trained and improved, forming a closed loop that guides subsequent rounds of autonomous experimentation. In this regard, high-throughput DFT can be further accelerated using ML to acquire sufficient training data [151]. Moreover, with the emergent quantum machine learning [152], it is hoped that the generative deep learning approaches can be further accelerated to explore a more comprehensive latent space. It is also noted that the generic inverse problem can all be tackled using such approaches, including the classical modelling such as classical DFT theory [153]. Therefore, we believe the potential for the generative deep learning approaches to revolutionize the field of materials discovery and engineering is enormous and disruptive, as they explore the latent space beyond the known design space, hence facilitating the development and employment of novel materials with unprecedented properties for various applications.

**Acknowledgements**

This work is funded by the Deutsche Forschungsgemeinschaft (DFG, German Research Foundation) - Project-ID 405553726 - TRR 270 and Project-ID 443703006 - CRC 1487 Iron, upgraded. The authors gratefully acknowledge the computing time provided to them on the high-performance computer Lichtenberg at the NHR Centers NHR4CES at TU Darmstadt.